**Gate Tunable Photovoltaic Effect in MoS$_2$ vertical P-N Homostructures**

*Simon A. Svatek*[1,2]*, *Elisa Antolin*[3], *Der-Yuh Lin*[4], *Riccardo Frisenda*[1], *Christoph Reuter*[5], *Aday J. Molina-Mendoza*[2], *Manuel Muñoz*[6], *Nicolás Agraït*[1,2], *Tsung-Shine Ko*[4], *David Perez de Lara*[1], *Andres Castellanos-Gomez*[1]*

[1] Instituto Madrileño de Estudios Avanzados en Nanociencia (IMDEA-Nanociencia), Faraday 9, Ciudad Universitaria de Cantoblanco, 28049, Madrid, Spain

[2] Departamento de Física de la Materia Condensada and Condensed Matter Physics Center (IFIMAC), Facultad de Ciencias, C/ Francisco Tomás y Valiente 7, Universidad Autónoma de Madrid, 28049, Madrid, Spain

[3] Instituto de Energía Solar, Universidad Politécnica de Madrid, E.T.S.I. Telecomunicación, Ciudad Universitaria s/n Madrid, Madrid 28040, Spain

[4] National Changhua University of Education, Bao-Shan Campus, No. 2, Shi-Da Rd, Changhua City 500, Taiwan, R.O.C

[5] Department of Electrical Engineering and Information Technology, Technische Universität Ilmenau, Gustav-Kirchhoff-Str. 1, Ilmenau 98693, Germany

[6] Instituto de Microelectrónica de Madrid (CNM, CSIC), C/ Isaac Newton 8, 28760 Tres Cantos, Madrid, Spain

*e-mail: simon.svatek@imdea.org and andres.castellanos@imdea.org

KEYWORDS: transition-metal dichalcogenides, photovoltaics, van der Waals assembly, p-type MoS2

**Abstract**: P-n junctions based on vertically stacked single or few layer transition metal dichalcogenides (TMDCs) have attracted substantial scientific interest. Due to the propensity of TMDCs to show exclusively one type of conductivity, n- or p-type, heterojunctions of different materials are typically fabricated to produce diode-like current rectification and photovoltaic response. Recently, artificial, stable and substitutional doping of $MoS_2$ into n- and p-type has been demonstrated. $MoS_2$ is an interesting material to use for optoelectronic applications due to the potential of low-cost production in large quantities, strong light-matter interactions and chemical stability. Here we report the characterization of the optoelectronic properties of vertical homojunctions made by stacking few-layer flakes of $MoS_2$:Fe (n-type) and $MoS_2$:Nb (p-type). The junctions exhibit a peak external quantum efficiency of 4.7 %, a maximum open circuit voltage of 0.51 V, they are stable in air and their rectification characteristics and photovoltaic response are in excellent agreement to the Shockley diode model. The gate-tunability of the maximum output power, the ideality factor and the shunt resistance indicate that the dark current is dominated by trap-assisted recombination and that the photocurrent collection depends strongly on the spatial extent of the space charge region. We demonstrate a response time faster than 80 ms and highlight the potential to integrate such devices into quasi-transparent and flexible optoelectronics.

**Introduction**: Two-dimensional layered materials offer a broad variety of building blocks with remarkable electronic, photonic and mechanical properties.[1] In particular, TMDCs, such as $WSe_2$ and $MoS_2$, are considered promising candidates for photovoltaic devices due to their strong light-matter interactions[2–4] and tunable (thickness-dependent) band gaps.[5,6] There is a growing number of fundamental[7–10] and applied[11–13] studies on thin layers of $MoS_2$. However, the formation of photovoltaic devices exclusively made of $MoS_2$ is hampered by the lack of samples with both transport characteristics, n-type and p-type, since $MoS_2$ typically exhibits only n-type doping. Previous experimental efforts to form junctions have therefore focused on layered heterojunctions formed by vertical van der Waals stacking of dissimilar materials, mostly in combination with $WSe_2$[14–17], and less frequently with other compounds such as GaTe[18], black phosphorus[19], and graphene.[20,21] Besides, the possibility to form homojunctions from $MoS_2$ has been explored with promising results but this could only be realised through superficial doping by exposing devices after processing to $AuCl_3$[22,23], through Schottky junctions[24], or through local electrostatic gating.[25–28] Recently, stable p-type conduction by substitutional niobium doping has been demonstrated in $MoS_2$[29], which enables the formation of all-$MoS_2$ p-n junctions. Here we demonstrate that, fabricating $MoS_2$ homostructures from vertical van der Waals assembly of p-type $MoS_2$ (0.5% Nb doping, ~3 × $10^{19}$ $cm^{-3}$) and n-type $MoS_2$ (either 0.5% Fe doping, or native n-type material), we can create solar harvesting p-n junctions. By fitting the current-voltage characteristics of these junctions with the Shockley diode model at various gate voltages, we extract all diode parameters and analyse in detail their gate-dependence.

The underlying physical mechanisms of rectification and photovoltaic response in vertical TMDC heterojunctions are fundamentally different to the mechanisms governing conventional p-n junctions based on the concept of selective contacts. It has been proposed[30] and experimentally shown[17,31] that in monolayer $MoS_2/WS_2$ heterojunctions the photovoltaic response and current-voltage characteristics are governed by interlayer tunnelling recombination of majority carriers and do not follow an exponential law.[17] The rectification is a result of the type-II band alignment with given band offsets for electrons and for holes. The maximum achievable photovoltage is limited by the smallest of these offsets, which in the case of MoS$_2$/WSe$_2$ is the band offset for electrons, approximately 0.7 eV.[30] Contrarily, in conventional p-n junctions selective contacts for electrons and holes are produced through p- and n-doping and the electrical characteristics are determined by drift and diffusion processes of minority carriers. The photovoltage is limited by the built-in field which in turn is limited by the material bandgap, which typically exceeds the band offsets in heterojunctions. In the case of bulk MoS$_2$ the direct band-gap is 1.8 eV and the indirect band-gap is 1.3 eV.[6] This consideration is crucial when assessing the potential of TMDC-based devices for solar applications and it emphasizes the need to explore vertical devices comprising selective contacts produced by stable atomic doping of TMDCs. In this work we present the characterization of vertical MoS$_2$ p-n homojunctions in the intermediate thickness regime (15 - 19 nm) between the monolayer and the bulk. We demonstrate that the current-voltage characteristics are in excellent agreement with the Shockley diode model, implying that the carrier dynamics in our MoS$_2$ homojunctions is predominantly governed by drift and diffusion processes. Ideality factors around 2.5 indicate trap-assisted recombination in the depleted regions. The performance of the devices is substantially affected by the gate voltage ($V_G$) through the modulation of the carrier densities and the electric field at the junction. In particular, a positive effect of $V_G$ on the open-circuit voltage ($V_{OC}$) is always accompanied by a negative effect on the short-circuit current ($I_{SC}$) which points to a carrier collection influenced by the spatial extent of the space charge region. Interestingly, not only the ideality factor and the reverse saturation current but also the parasitic shunt resistance is modulated by the gate, suggesting that tunnel-assisted recombination is still relevant in devices of moderate thickness (~ 15 nm).

**Results and Discussion:** The MoS$_2$:Nb and MoS$_2$:Fe single crystals were grown by the chemical vapour transport method described by Wang, S. Y. et al.[32] and Suh, J. et al..[29] (For an exhaustive study of the material composition see reference 29). Thin flakes from about 10 to 10³ µm² in size were prepared by micromechanical exfoliation of bulk crystals and deposited onto a viscoelastic stamp (GelFilm by GelPak)[33]. Upon optical inspection a flake was transferred onto an Au/Ti (40 nm/10 nm) metal lead on a 295 nm SiO$_2$/Si substrate by pressing the viscoelastic stamp onto the substrate and releasing it slowly. Figure 1a and b show schematically the fabrication steps along with optical images of the device. First, the n-type flake was deposited onto the substrate, partially overlapping with an Au/Ti (40 nm/10 nm) metal contact. Subsequently a p-type flake was transferred using the same method, overlapping with the first flake to form the vertical junction. The thicknesses of the flakes were determined to be 9 nm (MoS$_2$:Nb) and 6 nm (MoS$_2$:Fe) by atomic force microscopy.

We recorded the current density ($J$) as a function of the source drain voltage ($V_{SD}$) and the $V_G$. All measurements have been performed in ambient conditions. Figure 1c shows $J$-$V_{SD}$ characteristics of the p-n (MoS$_2$:Nb – MoS$_2$:Fe) homojunction in the dark and under illumination with monochromatic light (λ = 660 nm, intensity 80 mW/cm²) at various $V_G$ values. We observe rectifying $J$-$V_{SD}$ characteristics in the dark and photovoltaic behaviour under illumination. The photoresponse characteristics are modulated by the Si-gate electrode by changing the charge carrier density. The generated electrical power density $P_{el} = J V_{SD}$ for various $V_G$ is shown in Figure 1d. A maximum of $P_{el}$ = 0.42 mW/cm² and V$_{OC}$ = 0.51 V is generated at $V_G$ = -30 V. This exceeds reported $V_{OC}$s at similar intensities of van der Waals heterojunctions made of MoS$_2$/WSe$_2$[9,17], MoS$_2$/GaTe[18], and MoS$_2$/black phosphorous[19], although has been exceeded by graphene/MoS$_2$/WSe$_2$/graphene junction[17] due to enhanced extraction. Each $J$-$V_{SD}$ trace in Figure 1c has been fitted independently using a circuital model. It comprises a Shockley-diode and parasitic resistances in series ($R_S$) and in parallel ($R_P$) and it is described by the equation: $J = J_0 \left( \exp\left( \frac{e(V_{SD} - JR_S)}{nkT} \right) - 1 \right) + \frac{V_{SD} - JR_S}{R_P} - J_L$, where $k$ is the Boltzmann constant, e is the electron charge, $T$ is the temperature, $n$ is the ideality factor, $J_0$ is the reverse saturation current density of the diode and $J_L$ is the photogenerated current density. The fits are overlaid in Figure 1c as solid lines, showing strong agreement between the experimental data and the Shockley model. The parameters $n$, $J_0$, $J_L$, $R_S$, and $R_P$ that optimize each fitting are collectively plotted in Figure 2, and behave as continuous functions of $V_G$.

The agreement between the data and the Shockley-diode model implies that the carrier dynamics can be explained by the presence of a built-in field $E_{built-in}$ and the associated space charge region and neutral regions. This is illustrated in an ideal band diagram, see Figure 1e, in which the space charge region is represented as the grey area. Note that only the junction area, highlighted with dotted lines in Figure 1b, is considered in the band diagram. In our devices, drift and diffusion of minority carriers exclusively takes place in the junction area and the space charge region does not spread horizontally beyond. This implies that none of the observed effects originates from parasitic Schottky junctions at the contacts (further discussion is provided below regarding Figure 6).

The presence of a gate voltage ($V_G \neq 0$ V) produces either depletion or accumulation in the carrier population of the flakes and causes an additional field $E_{Gate}$ parallel to the built-in field $E_{built-in}$ at the junction (Figure 1f-g), which alters the band bending and the relative position of the bands with respect to the Fermi level. For negative $V_G$, the n-MoS$_2$ tends to be depleted, the p-MoS$_2$ accumulates carriers, and $E_{built-in}$ is reduced by $E_{Gate}$. This is accompanied by a reduction of the thickness of the space charge region since the density of ionized dopant atoms which provide the field is predetermined by the doping. A positive $V_G$ has the opposite effects. Thus, the thickness of the space charge region grows with $V_G$. The band bending introduced by $E_{Gate}$ resembles a forward source-drain bias for negative $V_G$ and vice versa. In our device the p-MoS$_2$ is expected to be degenerate[29], which implies that the carrier accumulation or depletion takes place mostly in the n-MoS$_2$. This has been taken into account in Figure 1f-g, where the Fermi-level appears pinned to the valence band in the p-type region.

The modulation of the thickness of the space charge region by $V_G$ affects the $J$-$V_{SD}$ characteristics plotted in Figure 1c and thereby the fitting parameters from the Shockley-diode model, which are presented in Figure 2. The Supplementary Information contains gate-dependent $J$-$V_{SD}$ characteristics in the dark (Figure S1). Ideality factor values in the range $n$ = [2 – 3.5] are indicative of recombination mechanisms that become less efficient as $V_{SD}$ increases. Such recombination is associated with coupled defects and possibly tunnelling mechanisms[34,35], which most likely occur in the space charge region where defects are not saturated and where the electric field can enhance tunnelling processes. The dramatic increase of $n$ with decreasing $V_G$ in our devices is consistent with tunnelling–assisted recombination at traps in the space charge region since tunnelling becomes less effective as the total field in the junction is reduced. It has to be noted that the depleted regions generated by $V_G$ at the MoS$_2$/SiO$_2$ interface do not necessarily contribute to this recombination because carriers are prevented to approach the MoS$_2$/SiO$_2$ interface by the presence of the depleting fields. The amount of recombination, characterized by $J_0$, increases with $V_G$ and thus, with the thickness of the space charge region. The dark current, open-circuit voltage and power densities improve with decreasing $V_G$, mostly due to the increase of $n$ and the decrease of $J_0$.

The parameter $J_L$ is the photogenerated current when no voltage is applied across the junction. Under those circumstances there can be no quasi-Fermi level split in the space charge region and therefore recombination can only take place in the rest of the device. Hence, $J_L$ accounts for the total photogenerated electron-hole pairs reduced by the recombination that takes place in the volume which is not occupied by the space charge region under short-circuit conditions.[36] Consequently, $J_L$ increases with $V_G$. The absolute value of $J_L$ in our devices is in the range [1.9 – 3.0 mA/cm²].

Interestingly, the fittings yield different $R_P$ values in the dark and under illumination. In the dark $R_P$ shows a strong dependence on $V_G$. $R_P$ quantifies local shuntings in the space charge region caused by tunnelling processes at coupled defects which is a known mechanism in conventional solar cells with high trap densities and $n >2$.[35] As above, the tunnelling is enhanced when the total field is increased by $V_G$, which leads to a decrease in $R_P$. Under illumination we find a constant $R_P$ independently of the gate. This is an apparent $R_P$ related to a modulation of the photogenerated current and not to shunting. It reflects an increase of the photogenerated current as $V_{SD}$ decreases which occurs due to the enlargement of the space charge region (similar to the case of positive $V_G$ discussed above). This is known from other technologies of very thin solar cells, such as CdTe and chalcopyrite-based devices, in which the space charge region occupies a significant part of the total volume, and it is known as voltage dependent collection.[37]

The series resistance $R_S$ of the flakes depends on the occupation levels. As VG increases the Fermi level is shifted towards the conduction band causing the resistance of the n-type (or p-type) flake to decrease (or increase). In the p-n junction we observe a decrease of the resistance with an increasing VG which implies that the gate-dependence of the n-type flake is predominant. This can be attributed to degenerate doping of the p-flake, to the lower thickness of the n-type flake and a gate-screening effect, due to which the p-type flake is not effectively controlled by $V_G$ since the n-type flake is screening the electric field. $R_S$ shows the same gate-dependence in the dark and under illumination. When illuminated $R_S$ is smaller than in the dark because of a higher over-all population. The values are in the range [60 - 100 MΩ] under illumination and [280 - 330 MΩ] in the dark.

Figure 1d shows that the positive effect of $V_G$ on the photogenerated current is counteracted by the opposite effect on the photovoltage, causing $P_{el}$ to increase as $V_G$ decreases. The previous analysis allows us to attribute both effects to the modulation of the space charge region. This result differs fundamentally from the behaviour of heterojunction-based WSe$_2$/MoS$_2$ devices, where a variation in $V_G$ affects directly the amount of interlayer recombination and leads to either an increase of photovoltage and photocurrent, or the reduction of both.[17] Note that the device performance does not improve indefinitely since $V_G$ introduces non-linear effects in the series resistance of the flakes leading to the junction. Such behaviour is more pronounced in a $MoS_2$:Nb – native $MoS_2$ junction, see Supplementary Information, Figure S3.

Figure 3a and b show the dependence of the $J$-$V_{SD}$ characteristics and $P_{el}$ on the incident power with above-bandgap illumination (λ = 660 nm). The electrical power density reaches a maximum of 0.68 mW/cm². The maximum in short-circuit current is $J_{SC}$ = 3.3 mA/cm² which corresponds to an external quantum efficiency of $EQE$ = 4.7 % and a responsivity of 25 mA/W, which is similar to what has been achieved in vacuum in $MoS_2$ homojunctions made by surface doping (30 mA/W).[22] A fill factor ($FF$) of 0.46 has been determined from max($P_{el}$)= $V_{OC}$ $J_{SC}$ $FF$. The dependences of $V_{OC}$ (Figure 3c, left axis, blue squares) and $J_{SC}$ (right axis, green circles) are, respectively, logarithmic and linear on the incident power which demonstrates that the $J$-$V_{SD}$ characteristics are dominated by the Shockley-diode despite the presence of parasitic resistances. Besides, this confirms that the photocurrent is predominantly caused by the photovoltaic effect as opposed to thermoelectric effects.[28]

By recording the photoresponse for various illumination wavelengths we characterize the junction further. Figure 4a shows $J$-$V_{SD}$ characteristics for wavelengths between 405 nm and 1050 nm. The response has a peak in the vicinity of the exciton energies 659 nm and 611 nm (in monolayer material)[38], where the optical absorption is enhanced. For photons with higher wavelengths, λ ≥ 850 nm, the device shows a negligible response, which is due to the reduced absorption of the indirect bandgap transitions. The response for wavelengths between 455 nm and 660 nm corresponds to $EQEs$ which range from 1.6 % to 4.5 %.

To determine the switching speed, we recorded the $J$ against the time while the illumination (λ = 660 nm, intensity 80 mW/cm²) was alternatingly turned on and off (Figure 4b). When the device is unbiased (black curve) we extract an upper limit of 80 ms for both the rise and decay times, which are defined as the time required for $J$ to rise (fall) to 90 % (10 %) of the difference between the initial and final value. Note, that this value is limited by our experimental setup. This temporal response is one order of magnitude faster than previously reported $MoS_2$ photodetectors with planar geometries[39], however, 9 orders of magnitude slower than values reported in graphene/TMD/graphene devices.[40] The strong differences in response times can be attributed to differences in the device geometries, carrier mobilities, flake thicknesses and extraction efficiencies. Response times in planar $MoS_2$ photodetectors are limited by the capacitance that is induced by trapped charge carriers. Faster response times in few-layer $MoS_2$ in ambient conditions (~10 ms) could to our knowledge only be realised by surface treatments, such as encapsulation with $HfO_2$ [41]. Under forward bias, $V_{SD}$ = 0.5 V (red curve), the photoresponse comprises also a photoconductance mechanism and the response time has been determined as 180 ms. We conclude that the photovoltaic effect in this geometry is at least 100 ms faster than other reported values. In line with results discussed earlier, our devices show a similar behaviour as conventional, selectively doped p-n junctions, in which the response time is dominated by diffusion mechanisms if operated in photovoltaic mode and a forward bias introduces a drift current which increases the response time.[42]

We have also prepared a p-n (MoS$_2$:Nb – native MoS$_2$, 13 nm – 6 nm) homojunction and compared it to the p-n (MoS$_2$:Nb – MoS$_2$:Fe) above. We find the same qualitative dependences of $n$, $I_0$, $R_P$, and $R_S$ on the gate voltage. Under illumination with $\lambda$ = 660 nm and an intensity of 80 mW/cm, the maximum power $P_{el}$ = 0.04 mW/cm² is generated at $V_G$ = -10 V and at $V_{SD}$ = 0.09 V. We extract $V_{OC}$ = 0.14 V and $FF$ = 0.4. The $J_{SC}$ = 0.64 mA/cm² corresponds to $EQE$ = 2.6 % and a responsivity of 14 mA/W. We attribute the lower device performance to the lower doping level of the native MoS$_2$ in comparison to the intentionally doped MoS$_2$ (for additional data and a discussion of the doping levels, see Supporting Information).

The van der Waals assembly technique allows to easily form p-n junctions on virtually any substrate. Here, we explore the potential for transparent and flexible substrates. Figure 5a-d show optical micrographs of a p-n (MoS$_2$:Nb – MoS$_2$:Fe, 10 nm – 9 nm) homojunction on a polycarbonate substrate in reflection- and transmission-mode. We find that ~50 % of the incident white light is transmitted through the electrodes. The transmittance of the device area varies in the range of 70 ± 15 %, suggesting that our materials may be used in quasi-transparent applications. Figure 5e shows the $J$-$V_{SD}$ characteristics in dark and under illumination of this flexible device. From circuit modelling we find $n$ = 2.46, $I_0$ = 0.3 pA, $R_S$ = 0.7 GΩ, and negligible $R_P$ in the dark and $n$ = 2.84, $I_0$ = 5.8 pA, $R_S$ = 70 MΩ and $R_P$ = 7 GΩ under illumination with $\lambda$ = 660 nm at a power of 80 mW/cm². We extract the maximum power density $P_{el}$ = 0.11 mW/cm² at $V_{SD}$ = 0.21 V. The other figures of merit are $EQE$ = 1.5 %, and responsivity of 8 mA/W for $J_{SC}$ = 0.66 mA/cm², $V_{OC}$ = 0.32 V, and $FF$ = 0.51. After bending the substrate 50 times to a local radius of curvature of 2.5 cm (see Figure 5f) the $J$-$V_{SD}$ characteristics remain unaltered. The lower device performance as compared to devices on SiO$_2$/Si substrates may be attributed to the absence of a reflective background.

To provide further evidence that the photocurrent is produced only in the junction area we fabricated another device (Figure 6b) and recorded scanning photocurrent maps with the set-up sketched in Figure 6a.[43,44] The device is illuminated with a fiber-coupled LED ($\lambda$ = 505 nm, spot diameter is 8 µm) and moved with an xy-micrometer stage while simultaneously recording the intensity of the reflected light and the photocurrent. The spatial map of the reflected light intensity allows to determine the position of the focus[43]. The photocurrent maps in Figure 6d and 6e were acquired in short-circuit and reverse bias conditions and reveal that the photovoltaic response exclusively occurs when the overlap region is illuminated. Under forward bias above $V_{OC}$, the resistance is determined by the resistance of the p-type flake which changes due to photoconductivity, see Figure 6f.

**Summary:** In summary, homojunctions based on few-layer MoS$_2$ have been fabricated and show external quantum efficiencies at 660 nm for different devices between 1.5 % and 4.6 %. We find that devices in this intermediate thickness regime (15 - 19 nm) between monolayers and bulk samples behave according to the Shockley-diode model. We have thoroughly analysed the gating effects on the photoresponse and conclude that the thickness of the space charge region decreases with the gate voltage which is correlated with an increase in device performance. We have demonstrated a gate-induced shifting of the maximum power point, $J_{SC}$ and $V_{OC}$. The device shows fast response times compared with previous studies on MoS$_2$ layers, which indicates that a vertical design could pave the way to faster MoS$_2$-based photodetectors. We have shown that devices made from substitutionally doped n-type material show superior device characteristics than devices made from native MoS$_2$. The p-n junctions demonstrate the potential of doped MoS$_2$ for quasi-transparent optical components in light harvesting cells and nanoscale optoelectronics.

**SUPPORTING INFORMATION**

Gate dependent *J-V*$_{SD}$ characteristics in the dark (S1). *J-V*$_{SD}$ characteristics at higher bias voltages (S2). Extend of the space charge region. Additional data regarding the MoS$_2$:Nb – native MoS$_2$ junction (S3). Comparison of gate dependence between junctions and single flakes (S4). Additional IV-curves of device in Figure 6 (S5).

**AUTHOR CONTRIBUTIONS**

SA Svatek has fabricated the devices and carried out the electrical characterisation. E Antolin generated the software that facilitates the electrical characterisation. DY Lin and TS Ko have fabricated the material. AJ Molina-Mendoz and M Munoz have carried out the AFM characterisation. R Frisenda and C Reuter have carried out the photocurrent mapping. SA Svatek, E Antolin, N Agrait, D Perez de Lara, AC Castellanos-Gomez have analysed the data. AC Castellanos-Gomez has coordinated the project. All authors have edited the manuscript and contributed to discussions.

**ACKNOWLEDGMENTS**

S.A.S. and N.A. acknowledge funding by the European Commission through the FP7 ITN MOLESCO (Project Number 606728). E. A. gratefully acknowledges financial support from L'Oréal-UNESCO through the Women in Science program. A.J.M-M. acknowledges the financial support of MICCINN (Spain) through the scholarship BES-2012-057346. D.Y. and T.S. acknowledge the support of the Ministry of Science and Technology of the Republic of China under the MOST 104-2112-M-018-004 and 104-2221-E-018-017. D. P. is thankful for funding from the Spanish Ministry of Economy and Competitiveness through FIS2015-67367-C2-1-P. R.F. thanks the Netherlands Organisation for Scientific Research (NWO) for the financial support through the research programme Rubicon with project number 680-50-1515. AC-G. acknowledges support from the BBVA Foundation through the fellowship "I Convocatoria de Ayudas Fundacion BBVA a Investigadores, Innovadores y Creadores Culturales" ("Semiconductores ultradelgados: hacia la optoelectronica flexible"), from the MINECO (Ramón y Cajal 2014 program, RYC-2014-01406), from the MICINN (MAT2014-58399-JIN) and from the Comunidad de Madrid (MAD2D-CM Program (S2013/MIT-3007)), and the European Commission under the Graphene Flagship, contract CNECTICT-604391.

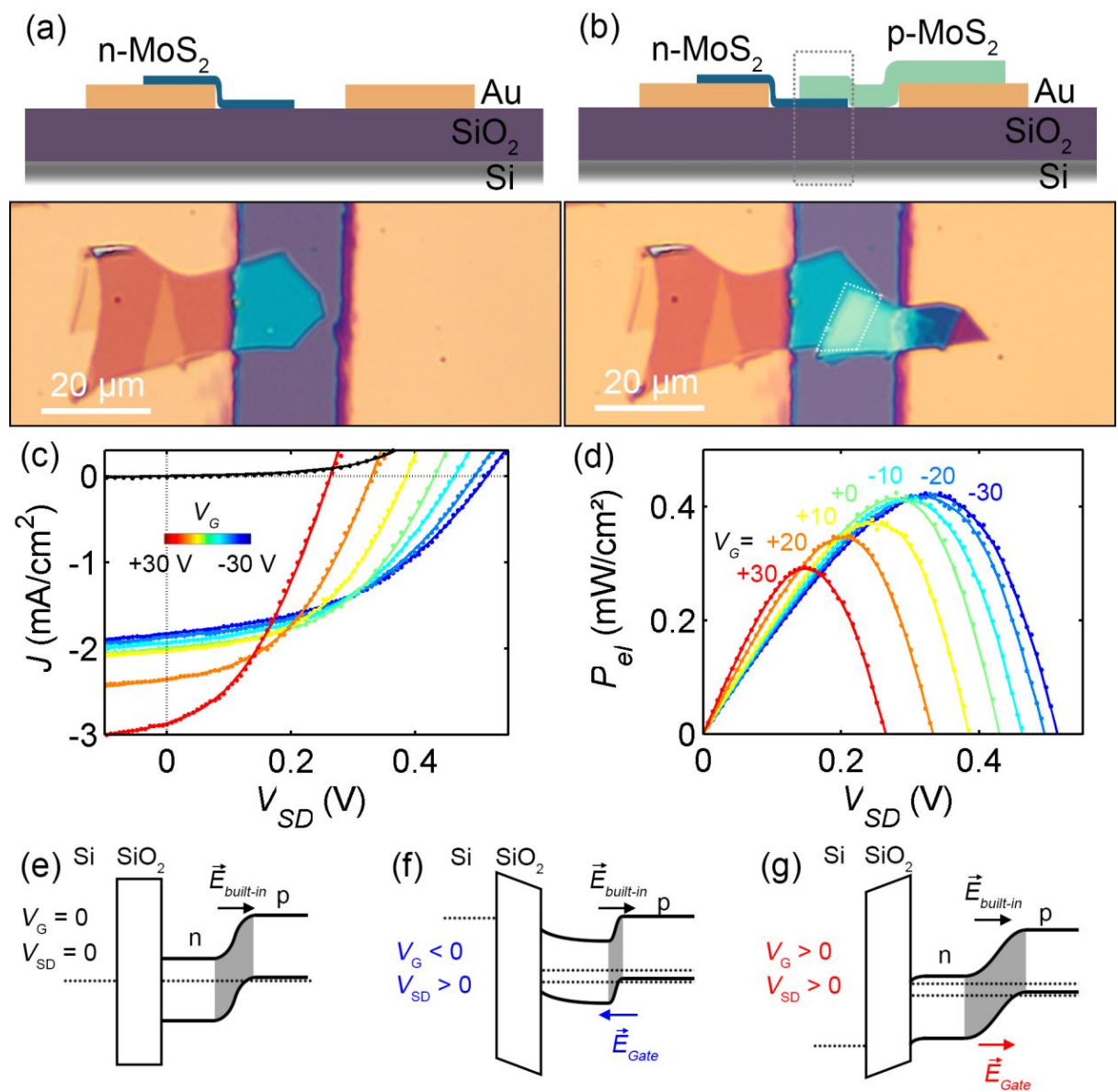

**Figure 1.** Device schematic and optical image of the n-type MoS$_2$ flake on top of Au/Ti lead before a) and after b) the transfer of p-type MoS$_2$. The dotted lines indicate the junction area. c) $J$-$V_{SD}$ curves taken under illumination with $\lambda$ = 660 nm at 80 mW/cm² at gate voltages from +30 V to -30 V in 10 V-steps (black: $J$-$V_{SD}$ curve in the dark at $V_G$ = 0 V). Experimental data is depicted as points and fits assuming the Shockley diode model are overlaid. d) The generated electrical power density $P_{el}$ against $V_{SD}$ at various $V_G$. (e-g) Band diagrams of the junction area at various values for $V_G$. The dotted lines represent the quasi Fermi levels for electrons and holes. The space charge region is highlighted in grey. $V_G$ introduces an electric field $E_{Gate}$ which enhances or reduces the total field and thereby contributes to the band bending.

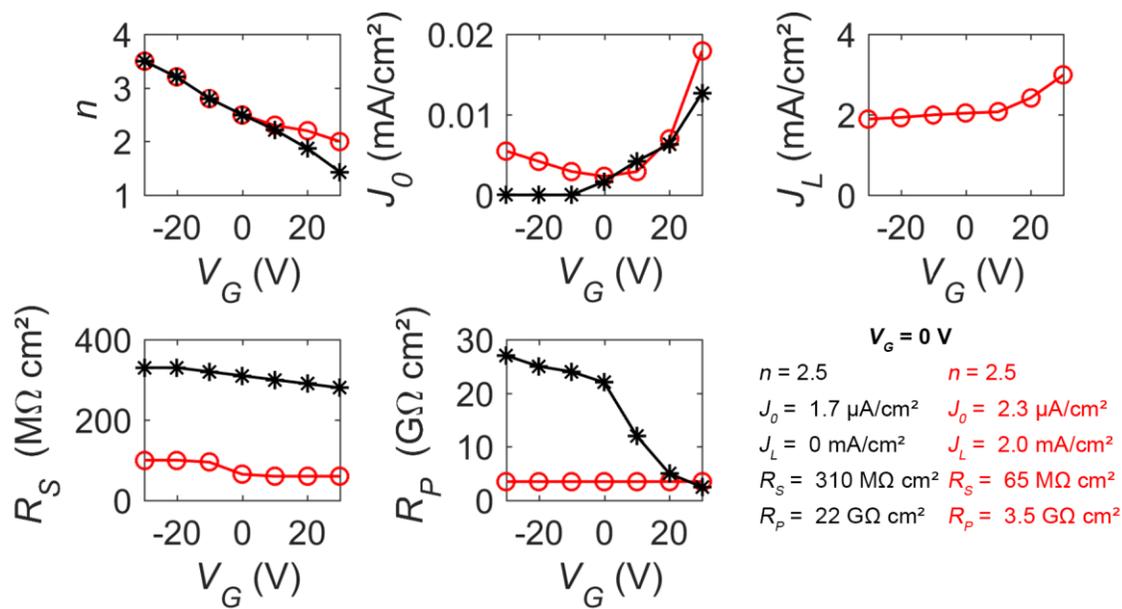

**Figure 2.** Fitting parameters extracted from Shockley-model-fits, see Figure 1 and S1. Red circles correspond to parameters under illumination (λ = 660 nm at 80 mW/cm²), black stars correspond to parameters in the dark.

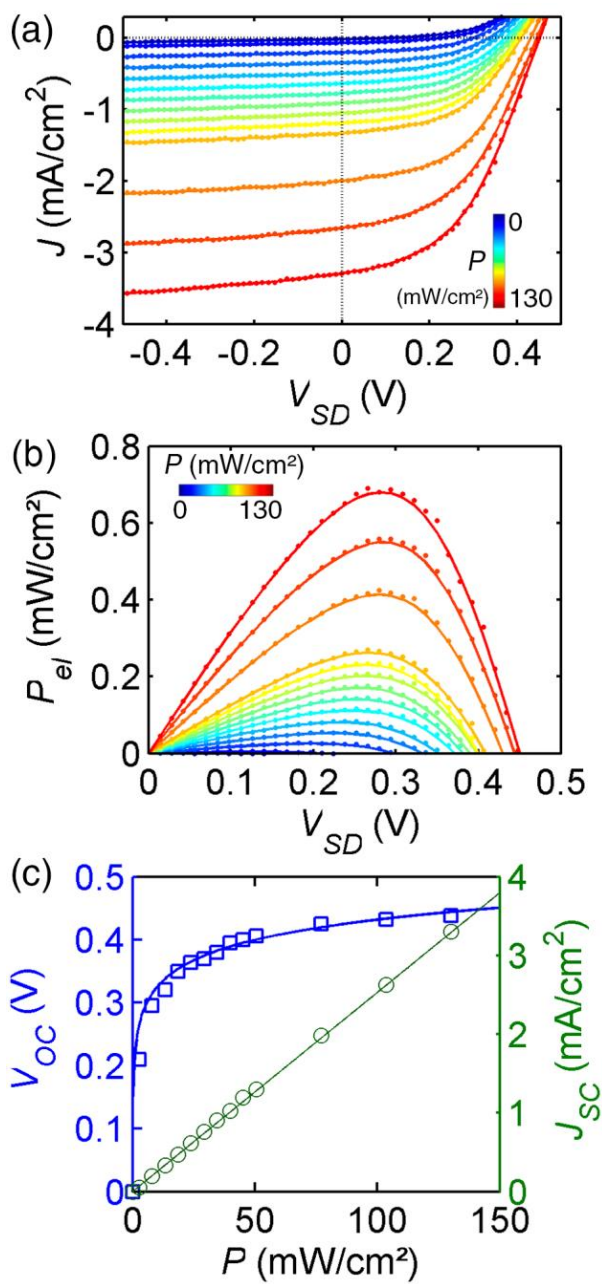

**Figure 3.** Device characteristics under illumination with λ = 660 nm at $V_G$ = 0 V. The incident power ranges from 0 to 130 mW/cm². a) Power dependence of $J$-$V_{SD}$ characteristics. b) Incident power dependence of the electrical power density $P_{el}$ vs. $V_{SD}$. c) Logarithmic (linear) power-dependence of $V_{OC}$ ($J_{SC}$) extracted from a).

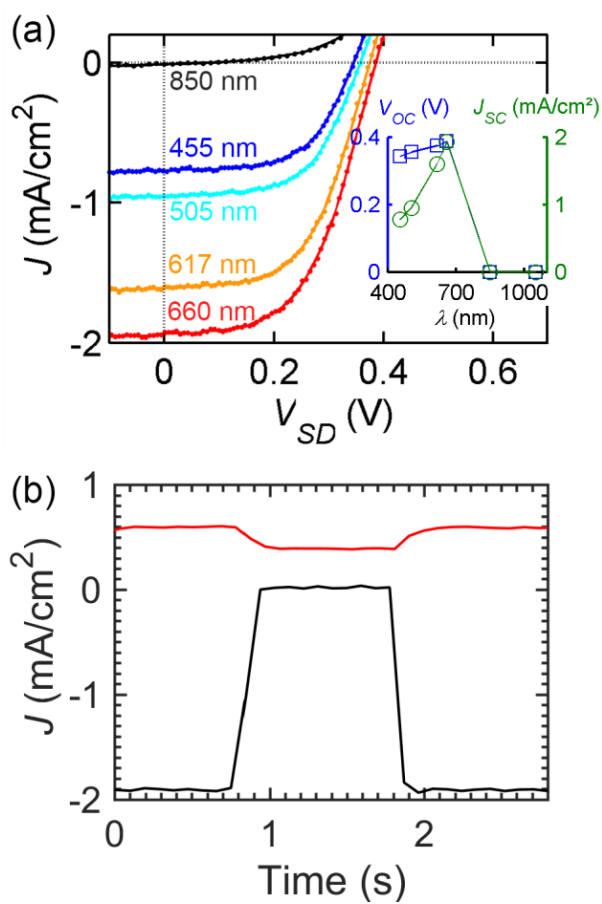

**Figure 4.** a) $J$-$V_{SD}$ curves at different illumination wavelengths at 80 mW/cm². The device shows no photoresponse at wavelengths λ ≥ 850 nm. The inset shows the dependence of $V_{OC}$ and $J_{SC}$ on the wavelength. b) Current density across the device against time at $V_G$ = 0 V. The illumination (80 mW/cm at λ = 660 nm) was alternatingly turned on and off. Under short circuit conditions (black) we extract 80 ms as an upper limit for the response time. The response time at at $V_{SD}$ = 0.5 V (red) is determined as 180 ms.

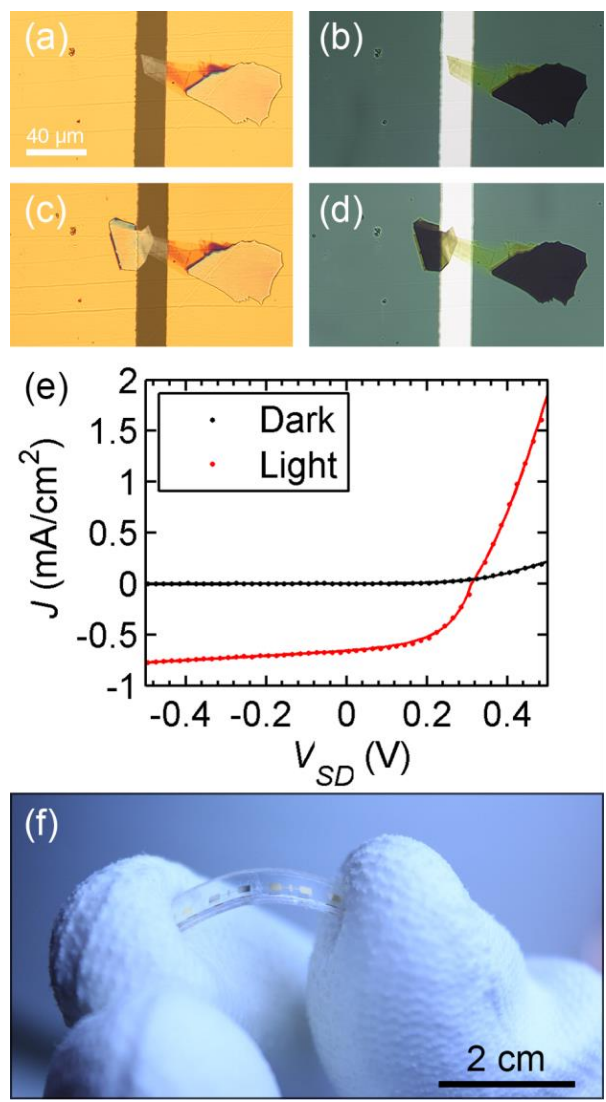

**Figure 5.** a)-d) Optical images of a quasi-transparent device with MoS$_2$:Nb (a, b) and the MoS$_2$:Nb - MoS$_2$:Fe junction (c, d) in bright field (a, c) and transmission-mode (b, d) on a polycarbonate substrate with Au/Ti (70 nm/15 nm) leads. e) *J-V$_{SD}$* curves in dark and under illumination with λ = 660 nm and 80 mW/cm². Fits assuming the Shockley diode model are overlaid as solid lines. f) A photograph of the device. The scale bar refers to the focal plane of the camera.

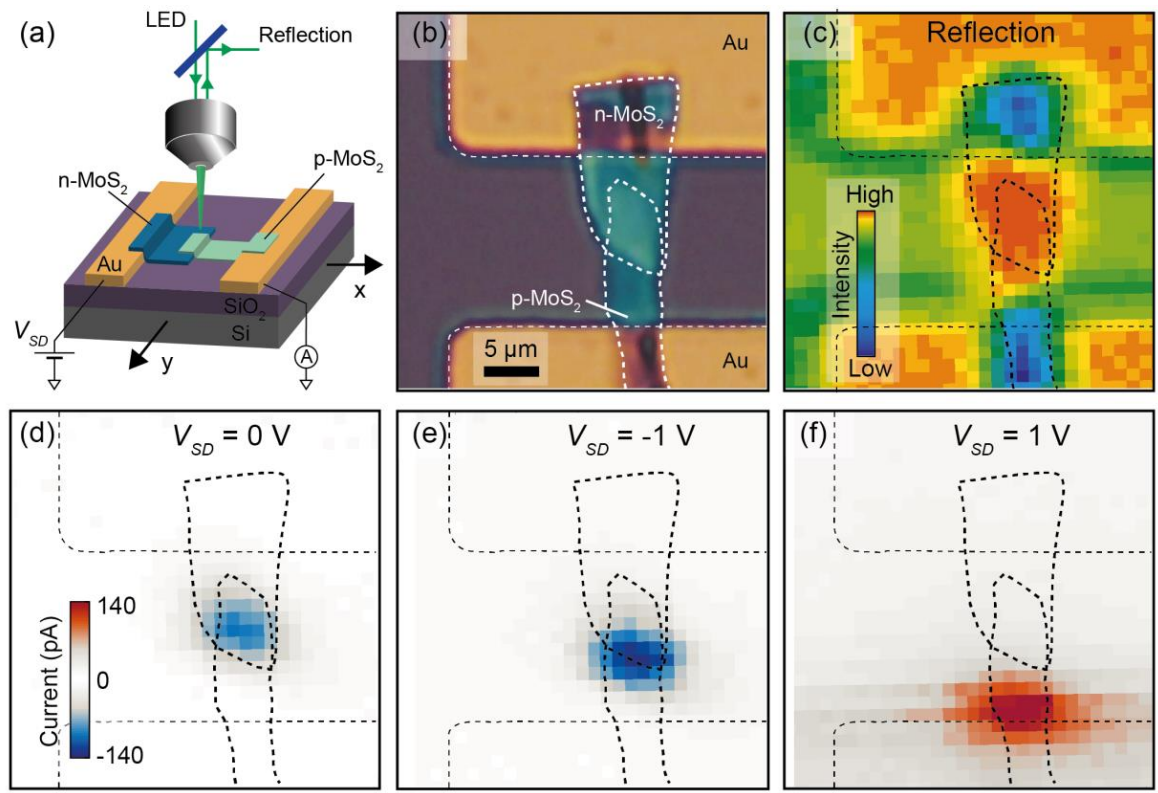

**Figure 6.** a) Sketch of the scanning photocurrent mapping set-up. A fibre-coupled LED ($\lambda$ = 505 nm, $P$ = 0.6 µW) illuminates the sample which is moved using an xy-micrometer stage. A reflection map allows to determine the position of the focus. By applying $V_{SD}$, and measuring the photocurrent with respect to the stage position the spatial origin of the photocurrent can be determined. b) Optical image of the MoS$_2$ p-n homojunction. The dashed lines indicate the position of the electrodes and the two flakes. c) Spatial map of the reflection intensity. d-f) Photocurrent maps recorded at $V_{SD}$ = 0 V, $V_{SD}$ = -1 V and $V_{SD}$ = 1 V. The photovoltaic response (in short-circuit and reverse bias conditions) originates in the overlap region. Under forward bias (above $V_{OC}$) the resistance of the device is determined by the resistance of the leads which change due to photoconductivity.

**Supporting information**

**Gate Tunable Photovoltaic Effect in MoS$_2$ vertical P-N Homostructures**

Simon A. Svatek[1,2]*, Elisa Antolin[3], Der-Yuh Lin[4], Riccardo Frisenda[1], Christoph Reuter[5], Aday J. Molina-Mendoza[2], Manuel Muñoz[6], Nicolás Agraït[1,2], Tsung-Shine Ko[4], David Perez de Lara[1], Andres Castellanos-Gomez[1]*

[1] Instituto Madrileño de Estudios Avanzados en Nanociencia (IMDEA-Nanociencia), Faraday 9, Ciudad Universitaria de Cantoblanco, 28049, Madrid, Spain

[2] Departamento de Física de la Materia Condensada and Condensed Matter Physics Center (IFIMAC), Facultad de Ciencias, C/ Francisco Tomás y Valiente 7, Universidad Autónoma de Madrid, 28049, Madrid, Spain

[3] Instituto de Energía Solar, Universidad Politécnica de Madrid, E.T.S.I. Telecomunicación, Ciudad Universitaria s/n Madrid, Madrid 28040, Spain

[4] National Changhua University of Education, Bao-Shan Campus, No. 2, Shi-Da Rd, Changhua City 500, Taiwan, R.O.C

[5] Department of Electrical Engineering and Information Technology, Technische Universität Ilmenau, Gustav-Kirchhoff-Str. 1, Ilmenau 98693, Germany

[6] Instituto de Microelectrónica de Madrid (CNM, CSIC), C/ Isaac Newton 8, 28760 Tres Cantos, Madrid, Spain

*e-mail: simon.svatek@imdea.org and andres.castellanos@imdea.org

**p-n junction ($MoS_2$:Nb – $MoS_2$:Fe): Gate dependent $J$-$V_{SD}$ characteristics in the dark**

The device shows gate-tunable rectification in the dark, see Figure S1. Since $V_G$ acts similar to an effective bias voltage in respect to the modulation of the barrier height and the extent of the space charge region, the device is closer to breakdown at high $V_G$. The breakdown voltage is therefore reduced by a positive $V_G$. The Shockley model equations do not model the reverse breakdown which accounts for the deviation of the model from the data in reverse bias.

The inset shows the gate leakage-current for this particular device. The leakage current has been determined for all devices. Devices with leakage currents above 150 pA at $V_G$ = 30 V or $V_{SD}$-dependent leakage were discarded.

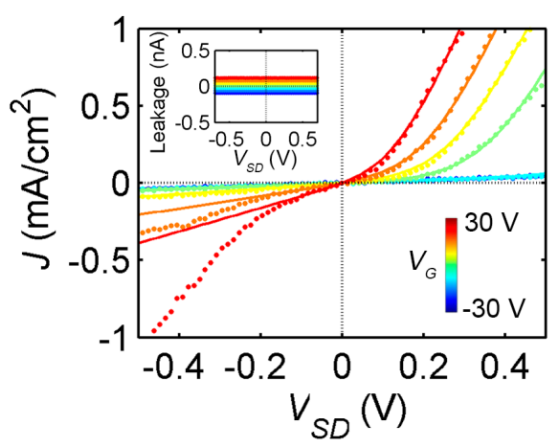

**Figure S1.** $J$-$V_{SD}$ curves taken in the dark at $V_G$ from +30 V to -30 V in 10-V steps. Experimental data is depicted as points and a Shockley-model-fit is overlaid. The inset shows the leakage-current to the gate electrode.

**p-n junction (MoS$_2$:Nb – MoS$_2$:Fe): $J$-$V_{SD}$ characteristics under illumination at high $V_{SD}$**

Figure S2 shows $J$-$V_{SD}$ curves at high $V_{SD}$. The current density $J$ saturates at high positive bias $V_{SD}$. We attribute this to either Schottky barriers formed at the interface between the flakes and the metal leads or barriers forming in the direction parallel to the flake when moving away from the junction area toward the electrode contacting the MoS$_2$.[1] However, most relevant for photovoltaic applications is quadrant IV in which the effects of the leads are small, as discussed in the main text regarding Figure 3(c).

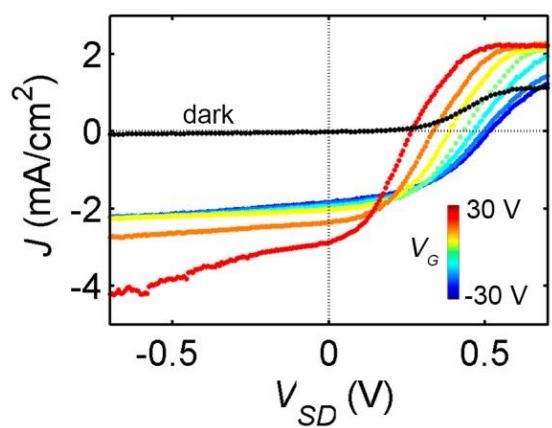

**Figure S2.** $J$-$V_{SD}$ curves taken under illumination with 660 nm at 80 mW/cm² at gate voltages from +30 V to -30 V in 10-V steps (black: $J$-$V_{SD}$ in the dark at $V_G$ = 0 V).

**The extent of the space charge region**

In order to estimate the potential of substitutional Nb doping to generate ultra-thin MoS2 homojunctions we can calculate the extent of the space charge region for the doping level achieved. The nominal doping level of the p-MoS2 is 0.5% Nb. The actual concentration of dopant atoms has been estimated to be 0.4% from electron spectroscopy for chemical analysis (ESCA) measurements. The bulk electrical doping extracted from sheet resistance measurements is 3.0 x 10$^{19}$cm$^{-3}$. Using the total depletion approximation[2] the thickness $x_p$ of the p-side of the p-n junction under equilibrium is given by

$$x_p = \sqrt{\frac{2\,\varepsilon_{MoS2}\varepsilon_0\,V_{built-in}N_D}{q\,N_A\,(N_A + N_D)}}$$

where $\varepsilon_{MoS2}$ is the relative permittivity of MoS$_2$ (11 from X. Chen et al.[3]), $\varepsilon_0$ is the permittivity of vacuum, q is the electron charge and $N_A$ ($N_D$) is the acceptor (donor) density in the p (n) flake. $V_{built-in}$ is the built-in potential which can be calculated as $V_{built-in} = kT\,Ln(N_A\,N_D/n_i^2)$ where $n_i$ is the intrinsic concentration (1.6 × 10$^8$ cm$^{-3}$, from S. Kim et al.[4]). As a limit, assuming that $N_D = N_A$, we obtain $x_p$ = 5 nm.

**p-n junction (MoS$_2$:Nb – native MoS$_2$)**

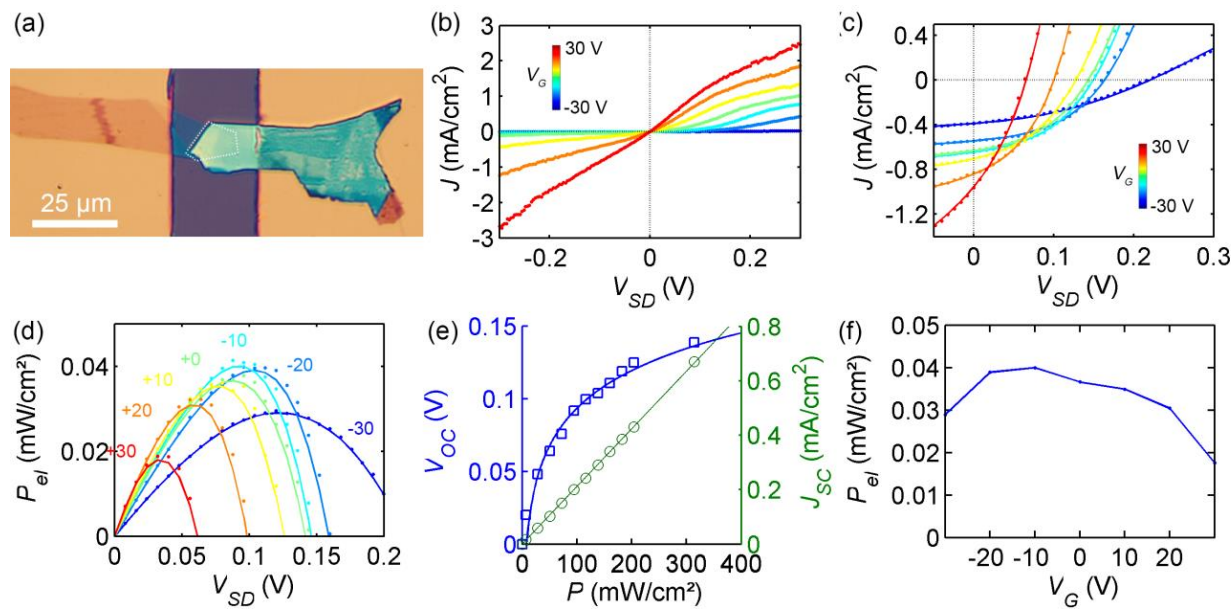

**Figure S3.** a) Optical image of the device. The dotted line indicates the junction area. b) $J$-$V_{SD}$ curves taken in the dark and c) under illumination ($\lambda$ = 660 nm, intensity 80 mW/cm²) with various $V_G$. Fits assuming a Shockley-diode are overlaid. d) Electrical power density $P_{el}$ as a function of $V_G$. e) Logarithmic (and linear) power-dependence of $V_{OC}$ (and $I_{SC}$) extracted from illumination power-dependent measurements. f) Gate-dependence of the electrical power $P_{el}$.

**Gate dependence of dark current. Comparison to flakes**

We have fabricated devices consisting of flakes of each of the three materials on 295 nm SiO$_2$/Si substrates with Au/Ti contacts. None of these devices exhibit a photovoltaic response. Therefore the photovoltaic response discussed in the main text originates indeed at the p-n junction and not at any other barrier, such as Schottky barriers at the contacts.

The gate dependences of the dark current in forward bias for single-flake devices are shown in figure S4. We find that the dark current across MoS$_2$:Nb decreases with $V_G$ and it increases across the MoS$_2$:Fe and the native MoS$_2$ with $V_G$, confirming p- and n-doping. In our p-n junctions, we find the same qualitative gate dependence as in the n-type flakes. This indicates that the total conductance of the p-n junction is mostly influenced by the n-type flakes presumably due to their lower thickness.

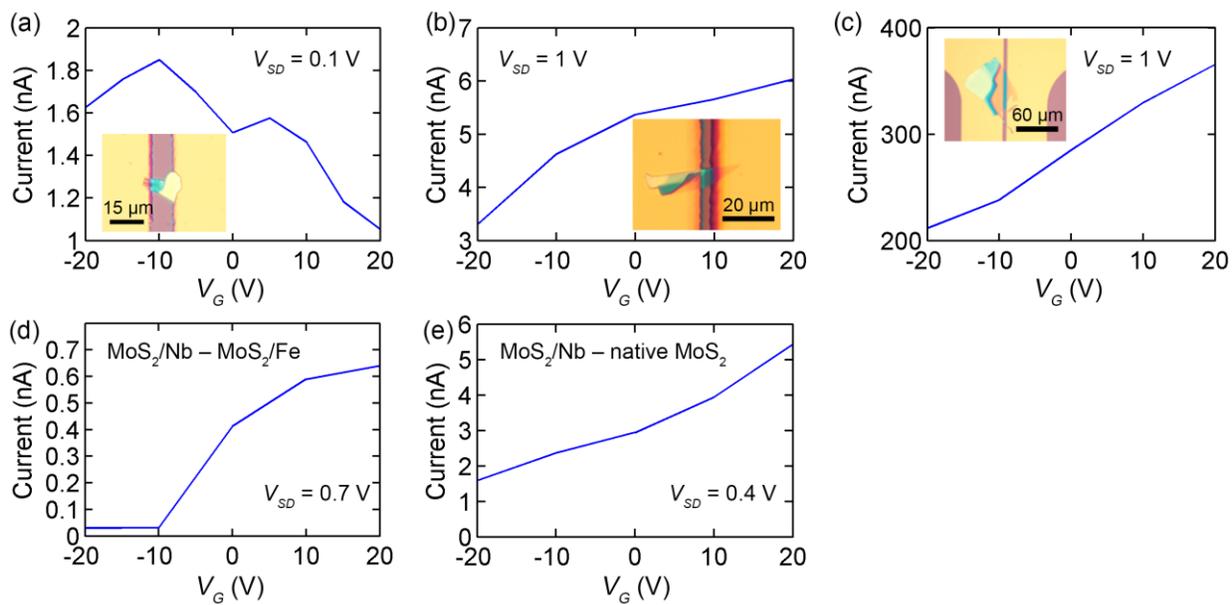

**Figure S4.** Gate dependence of dark current in forward bias conditions. a) MoS$_2$:Nb. b) MoS$_2$:Fe. c) Natural MoS$_2$. d) p-n junction, MoS$_2$:Nb.-MoS$_2$:Fe. e) p-n junction, MoS$_2$:Nb-natural MoS$_2$. The insets show optical images of the devices.

**Additional IV-curve**

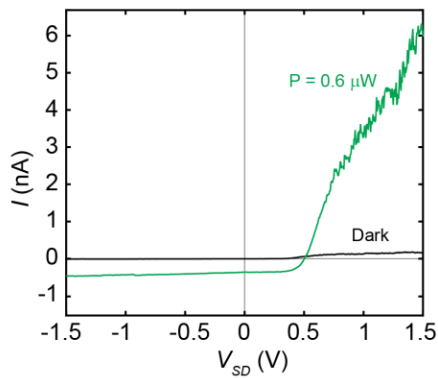

**Figure S5.** *I-V$_{SD}$* curves taken in the dark and under illumination (λ = 530 nm, intensity 6 µW) of device in Figure 6.

**SUPP. INFO. REFERENCES**